\pdfoutput=1

\documentclass[11pt]{article}

\usepackage[preprint]{acl}

\usepackage{times}
\usepackage{latexsym}

\usepackage[T1]{fontenc}

\usepackage[utf8]{inputenc}

\usepackage{microtype}

\usepackage{inconsolata}

\usepackage{graphicx}
\usepackage{float}
\usepackage{booktabs}
\usepackage{listings}
\title{Golden-Retriever: High-Fidelity Agentic Retrieval Augmented Generation for Industrial Knowledge Base}

\author{Zhiyu An$^{\sigma\mu 1}$\;\; Xianzhong Ding$^{\gamma}$\;\; Yen-Chun Fu$^\mu$\;\; Cheng-Chung Chu$^\mu$\;\; Yan Li$^\mu$\;\; Wan Du$^{\sigma}$ \\
  $\sigma:$ University of California, Merced, CA, USA \\
  $\mu:$ Western Digital Corporation, CA, USA \\
  $\gamma:$ Lawrence Berkeley National Laboratory, CA, USA \\
  \texttt{\{zan7, wdu3\}@ucmerced.edu} \\
  }

\begin{document}
\maketitle
\begin{abstract}
This paper introduces Golden-Retriever, designed to efficiently navigate vast industrial knowledge bases, overcoming challenges in traditional LLM fine-tuning and RAG frameworks with domain-specific jargon and context interpretation. Golden-Retriever incorporates a reflection-based question augmentation step before document retrieval, which involves identifying jargon, clarifying its meaning based on context, and augmenting the question accordingly. Specifically, our method extracts and lists all jargon and abbreviations in the input question, determines the context against a pre-defined list, and queries a jargon dictionary for extended definitions and descriptions. This comprehensive augmentation ensures the RAG framework retrieves the most relevant documents by providing clear context and resolving ambiguities, significantly improving retrieval accuracy. Evaluations using three open-source LLMs on a domain-specific question-answer dataset demonstrate Golden-Retriever’s superior performance, providing a robust solution for efficiently integrating and querying industrial knowledge bases.
\end{abstract}

\section{Introduction}\label{Sec: Introduction}

\footnotetext[1]{Work conducted while interning at Western Digital Corporation.}

\begin{figure*}[t]
  \includegraphics[width=\textwidth]{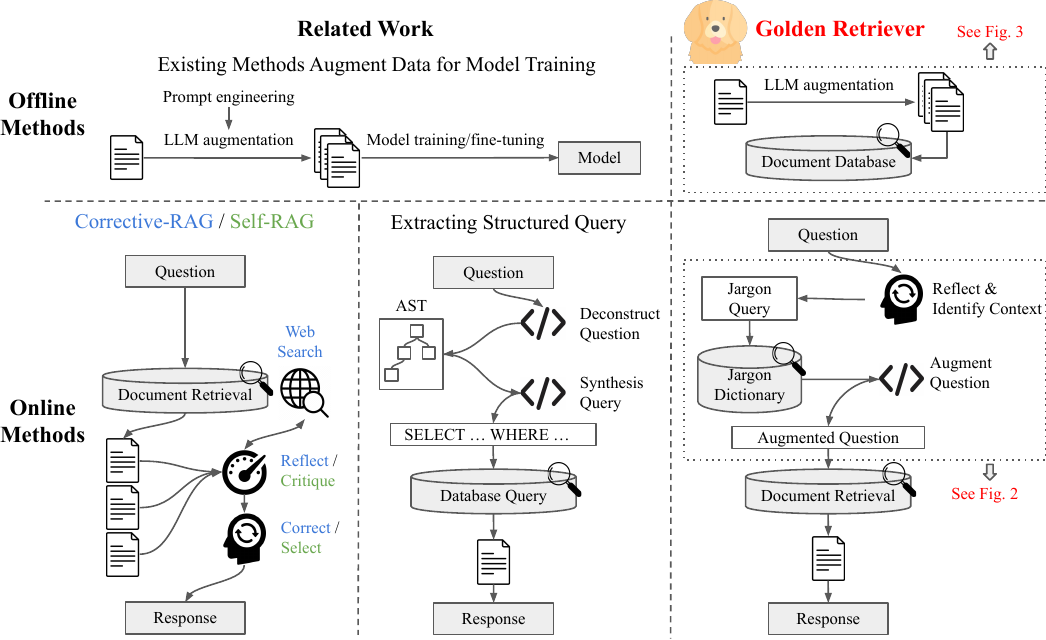}
  \caption{An illustration comparing our method with related works. We consider two types of methods: offline and online. On the upper-left, existing offline methods use LLMs to generate datasets for training. The upper-right shows our offline method, using LLMs to enhance the document database for the online phase. Online methods are depicted in the lower part of the figure. From lower-left to lower-right: Corrective RAG and Self-RAG modify the response of RAG after the document retrieval step. If the user's question is ambiguous or lacks context, RAG cannot retrieve the most relevant documents, limiting the effectiveness of these methods. Another approach deconstructs the question into an AST and synthesizes SQL queries accordingly, improving query fidelity but only applicable to SQL queries. Our method reflects upon the question, identifies its context, and augments the question by querying a jargon dictionary before document retrieval. The augmented question allows RAG to faithfully retrieve the most relevant documents despite ambiguous jargon or lack of explicit context.} \label{Fig: Main}
\end{figure*}

Technological companies maintain massive collections of proprietary documents generated over years, such as training materials, design documents, and research outputs. Engineers, especially new hires, are expected to quickly query these documents or assimilate the new knowledge in these documents. However, navigating in the large number of documents is challenging. These domain-specific documents normally have many abbreviations and jargons unique to their technical community, further complicating the problem.

Large Language Models (LLMs) offer excellent performance for general question-answering tasks \cite{petroni2019language, hu2021lora}. To make a pre-trained LLM incorporate a company's domain-specific knowledge, we may fine-tune it over the company's proprietary documents. However, fine-tuning is computationally expensive, generalize poorly to new knowledge due to the Reversal Curse \cite{berglund2023reversal}, and limited in capacity, as  it may overwrite old knowledge \cite{roberts2020much, zhai2024investigating}.

Retrieval Augmented Generation (RAG) \cite{lewis2020retrieval} offers a flexible and scalable approach for utilizing large document collections. RAG consists of an embedding model, a document database, and a LLM. During offline preparation, RAG embeds document chunks into the document database that retains semantic information. When an user asks a question, RAG first retrieves relevant document chunks according to semantic similarity. Then, the retrieved chunks are incorporated into prompts for the LLM, which then generates an answer. The output of RAG is the answer generated by the LLM based on the document chunks. This allows dynamic updates of knowledge base for an LLM without retraining it.

Despite its advantages, RAG also faces challenges to be used for domain-specific documents. 
First, since some jargons and abbreviations only appear in proprietary documents, RAG's LLM backbone may hallucinate and misinterpret them. 
Existing methods like Corrective RAG \cite{yan2024corrective} and Self-RAG \cite{asai2023self} enhance the LLM's response post-retrieval. But when user's question contains ambiguous jargons, RAG fails to retrieve the most relevant documents, limiting the effectiveness of post-retrieval enhancements.
To disambiguate user's question before retrieval, another approach \cite{kochedykov2023conversing} deconstructs vague questions into an Abstract Syntax Tree (AST) and synthesizes SQL queries, improving query fidelity. But their work is limited to SQL instead of natural language documents. 
Second, while identifying the context of the question is crucial for retrieving relevant documents, the actual questions asked by the user rarely contain context information.
The mentioned approach \cite{kochedykov2023conversing} trains a transformer-based text classifier to classify user's question to a predefined set of contexts, but this requires collecting a dedicated training dataset with diverse questions, which is exceedingly laborious.

To tackle these challenges, we propose Golden-Retriever, which enhances the traditional RAG framework with a reflection-based question augmentation step before document retrieval. Golden-Retriever identifies jargons and clarifies their meaning based on the context. By ensuring that the context is identified and ambiguities are resolved before document retrieval, Golden-Retriever significantly reduces the risk of misinterpretation and improves the relevance of retrieved documents.

Golden-Retriever includes both offline and online processes. The offline part involves a data pre-processing step where Optical Character Recognition (OCR) is used to extract text from various document formats. This text is then summarized and contextualized by LLMs to enhance the document database, ensuring that documents are more likely to be relevant when queried. Unlike existing offline methods that use LLMs for model training or fine-tuning to improve cross-language performance or generate counterfactual data \cite{whitehouse2023llm, sen2023people}, our approach focuses on augmenting the RAG document database directly.

The online part is an interactive process that occurs each time a user asks a question.
It starts with identifying jargons and context within the user's query using LLMs. The identified jargons is then queried against a jargons dictionary to retrieve accurate definitions and descriptions. This information is used to augment the original question, providing clear context and resolving any ambiguities. The augmented question is then used as input for the RAG framework, ensuring that the most relevant and accurate documents are retrieved. 

In summary, our contributions are as follows:
\begin{itemize}
    \item We identify the challenges of using LLMs for knowledge bases in real-world deployments.
    \item We propose Golden-Retriever, an agentic derivative of RAG featuring reflection-based question augmentation before document retrieval, enabling RAG to retrieve the most relevant documents despite ambiguous jargons and lack of context.
    \item We evaluate Golden-Retriever with three open-source LLMs and compare its performance with baselines on a dedicated, domain-specific question-answer dataset.
\end{itemize}

\section{Related Work}
Current RAG techniques often fall short of the ideal scenario for handling domain-specific queries in industrial knowledge bases. 

Vanilla RAG \cite{lewis2020retrieval}, for instance, struggles with accurately interpreting domain-specific jargons. When asked, \textit{"What is the PUC architecture of Samsung or Hynix NAND chip?"}, the system incorrectly interprets \textit{"PUC"} as \textit{"Process-Unit-Controller"} instead of the correct \textit{"Peripheral Under Cell"}. This misinterpretation highlights the problem of hallucination, where the model generates incorrect or nonsensical information based on ambiguous input. This issue is further illustrated in Figure \ref{Fig: Main}, which shows that both Corrective RAG \cite{yan2024corrective} and Self-RAG \cite{asai2023self} attempt to modify the response after the document retrieval step. However, if the initial retrieval is flawed due to misinterpreted jargons or lack of context, these post-processing techniques cannot fully rectify the inaccuracies.

Moreover, Corrective-RAG and Self-RAG focus on refining the generated responses after retrieval, which is inherently limited if the retrieved documents themselves are not relevant. As depicted in Figure \ref{Fig: Main}, these methods fail to address the root cause: the ambiguity in the user's question and the initial retrieval process. A related approach by \cite{kochedykov2023conversing} aims to address vague questions by deconstructing them into an AST and synthesizing SQL queries accordingly. While this method improves query fidelity, it is limited to SQL queries and does not generalize to broader question-answering scenarios. Figure \ref{Fig: Main} illustrates this limitation, showing that while the method can disambiguate and structure queries more effectively, it is not applicable to general retrieval tasks where context and jargon interpretation are crucial.


\begin{figure*}[t]
  \includegraphics[width=\textwidth]{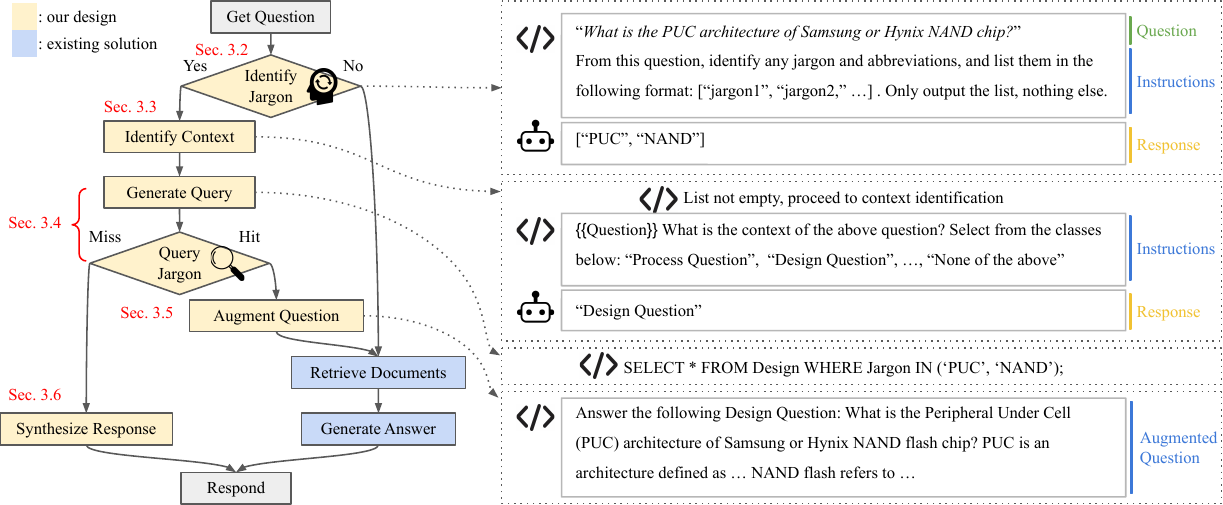}
  \caption{Left: the workflow diagram of the online inference part of Golden-Retriever. Right: example interactions between the system and the LLM at the intermediate steps of the workflow. The system prompts LLM to generate intermediate responses, which are saved, accessed, and used for future steps in the workflow.} \label{Fig: Framework}
\end{figure*}

\begin{figure}[t]
  \includegraphics[width=\columnwidth]{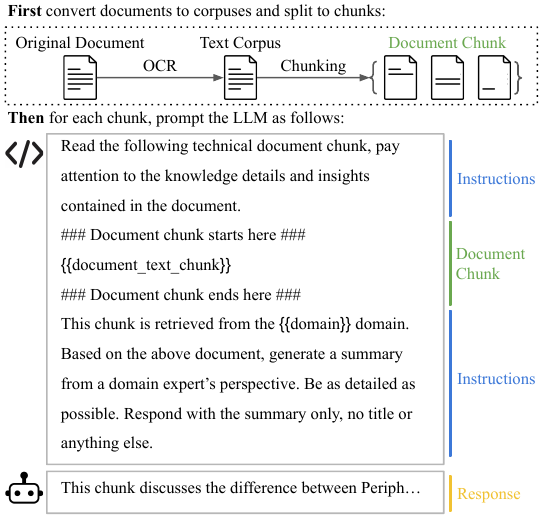}
  \caption{Section \ref{Sec: LLM-driven Document Augmentation}. Illustration of document pre-processing and an example prompt implementation of the LLM-Driven Document Augmentation Process.} \label{Fig: Data Augmentation}
\end{figure}

\section{Method}

Golden-Retriever consists of offline and online parts. The offline part is a data pre-processing step that occurs before the deployment of the knowledge base chatbot, described in Section \ref{Sec: LLM-driven Document Augmentation}. The online part is an interactive process that takes place every time a user asks a question, detailed in Sections \ref{Sec: Identify Jargon} through \ref{Sec: Response In Case of Query Miss}. 

\subsection{LLM-Driven Document Augmentation} \label{Sec: LLM-driven Document Augmentation}




The offline part of Golden-Retriever focuses on enhancing the document database to improve the relevance of retrieved documents. This process begins by collecting the company's original documents, such as slides, images with embedded text, and tables, to form the knowledge base. These documents are often varied in format and content, lacking a clear narrative, which can lead to low relevance scores when queried with RAG.

To address this, we use OCR to extract text from these documents and split it into smaller, manageable chunks for processing. For the Meta-Llama-3 model, these chunks are approximately 4,000 tokens each. Each chunk is then processed using an LLM to generate summaries from the perspective of a domain expert, leveraging the LLM's semantic understanding and in-context learning abilities. This augmented data is added to the document database, making it more likely to retrieve relevant documents when queried (Figure \ref{Fig: Data Augmentation}).

\subsection{Identify Jargons}\label{Sec: Identify Jargon}

The first step in the online process involves identifying jargons and abbreviations within the user's question. This step is essential because many domain-specific questions include specialized terms that require clarification to ensure accurate interpretation. To identify these terms, we utilize a prompt template designed to instruct the LLM to extract and list all jargons and abbreviations found in the input question. This process ensures that all potentially ambiguous terms are recognized, facilitating their resolution in later steps. The identified jargons and abbreviations are outputted in a structured format for further processing.

We choose to use the LLM for this task because traditional string-exact-match methods are inadequate. These methods may fail to detect jargons that are mistyped or not yet included in the dictionary, which could lead to misinterpretation in the following process. The LLM's ability to adapt to new terms provides a more robust solution. This step is represented as a two-way branching node in the workflow, shown in Figure \ref{Fig: Framework}. If the resulting list is empty, the main program proceeds along the "No" path; otherwise, it follows the "Yes" path. The structured response containing the identified terms is saved in a temporary file, which is then accessed by the main program to determine the next steps in the workflow.

\subsection{Identify Context} \label{Sec: Context Identification}

After identifying jargon, it is crucial to determine the context in which the question is asked, as the meaning of terms can vary significantly across different contexts. For instance, "RAG" could mean "Retrieval Augmented Generation" in the context of LLMs or "Recombination-Activating Gene" in genetics. To accurately interpret the context, we use a similar reflection step as in jargon identification. This involves designing a prompt template that takes the question as input. The prompt contains a list of pre-specified context names and their descriptions. The LLM uses this prompt to identify the context of the question. Few-shot examples with Chain-of-Thought (CoT) prompting are applied to enhance performance, guiding the LLM to respond in a specified data structure. The identified context is then stored and accessed by the main program for further processing.

Using simpler methods, such as transformer-based text classifiers like those used in \cite{kochedykov2023conversing} to classify user intent, would require a dedicated training dataset. This is impractical for our application due to the extensive effort and resources needed to create such a dataset. Instead, we opt for an "LLM as backend" approach, which, despite incurring higher computational costs, does not require a dedicated training dataset and can be run efficiently on a local server. By identifying the context before document retrieval, we ensure that the meaning of jargons and abbreviations is accurately interpreted, which is essential for retrieving the most relevant documents and providing accurate answers.

\subsection{Query Jargons} \label{Sec: Jargon Query}

\begin{table*}[t]
\caption{Question answering experiment results. We use quizzes from six different domains of the new-hire training documents for engineers as test questions. All questions are multiple choice questions. Average scores across five trials are shown. The best scores are in \textbf{bold}.}\label{Tab: experiment 1}
\begin{tabular}{r|ccc|ccc|ccc}
\toprule
             & \multicolumn{3}{c|}{Vanilla LLM} & \multicolumn{3}{c|}{RAG}                     & \multicolumn{3}{c}{Golden Retriever (Ours)}                     \\
             & Llama3    & Mistral    & Shisa   & Llama3        & Mistral       & Shisa        & Llama3        & Mistral       & Shisa        \\
\midrule
Quiz 1 - 10 Q & 3.2       & 4.0        & 4.0     & 5.0           & 3.0           & 3.0          & \textbf{6.0}  & 5.8           & 4.6          \\
Quiz 2 - 10 Q & 7.0       & 6.0        & 7.0     & \textbf{10.0} & \textbf{10.0} & 8.0          & \textbf{10.0} & \textbf{10.0} & 8.0          \\
Quiz 3 - 9 Q  & 4.2       & 5.0        & 5.0     & 6.0           & 7.0           & 4.0          & 7.0           & \textbf{8.0}  & 5.0          \\
Quiz 4 - 10 Q & 3.6       & 3.0        & 1.0     & 2.0           & 1.0           & 1.0          & \textbf{6.0}  & 4.0           & 4.0          \\
Quiz 5 - 10 Q & 1.2       & 4.0        & 2.0     & 1.0           & 3.0           & 2.0          & \textbf{5.0}  & 3.0           & \textbf{5.0} \\
Quiz 6 - 9 Q  & 2.0       & 1.0        & 2.0     & 3.0           & 3.0           & \textbf{4.0} & \textbf{4.0}  & 3.0           & \textbf{4.0} \\
\midrule
Total Score & 21.2 & 23.0 & 21.0 & 27.0 & 27.0 & 22.0 & \textbf{38.0} & 33.8 & 30.6 \\
\bottomrule
\end{tabular}
\end{table*}




Once the jargon and context have been identified, the next step is to query a jargon dictionary for extended definitions, descriptions, and notes on the identified terms. This step is essential for providing the LLM with accurate interpretations of the jargon, ensuring that the augmented question is clear and unambiguous.

This process involves querying a SQL database with the list of jargon terms identified in Section \ref{Sec: Identify Jargon}. The jargon list is inserted into a SQL query template, which is then processed to retrieve the relevant information from the jargon dictionary. The retrieved information includes extended names, detailed descriptions, and any pertinent notes about the jargon. We choose not to use the LLM to generate SQL queries directly, as described in \cite{qin2023toolllm} and \cite{li2024can}. Generating SQL queries with LLMs can introduce uncertainties regarding query quality and safety, and can also increase inference costs. Instead, by using a code-based approach to synthesize the SQL query, we ensure that the queries are verifiably safe and reliable.

The detailed information obtained from this step is crucial for augmenting the user's original question. It allows for accurate context and jargon interpretation, which is fundamental for the RAG process to retrieve the most relevant documents and generate precise answers.

\subsection{Augment Question}



With the jargon definitions and context identified, the next step is to augment the user's original question to include this additional information. This augmentation ensures that the RAG process retrieves the most relevant documents by providing clear context and resolving any ambiguities in the question. This step involves integrating the original question with the context information and the detailed jargon definitions obtained from Sections \ref{Sec: Context Identification} and \ref{Sec: Jargon Query}. The augmented question explicitly states the context and clarifies any ambiguous terms, facilitating enhanced document retrieval.

The process is automated, with the code taking the original question and the results from the context and jargon identification steps and combining them into a structured template. The context information grounds the LLM to the specified scenario, and the jargon definitions add relevant notes to clarify terms. The augmented question then replaces the user's original question and is used as input for the RAG framework, ensuring that the most relevant and accurate documents are retrieved.

\subsection{Query Miss Response} \label{Sec: Response In Case of Query Miss}



In some cases, the system may not find any relevant information for certain jargon terms in the dictionary. To handle such scenarios, Golden-Retriever has a fallback mechanism that synthesizes a response indicating that the database is unable to answer the question due to missing information. The system instructs the user to check the spelling of the jargon or contact the knowledge base manager to add new terms. This step ensures that the system maintains high fidelity and avoids generating incorrect or misleading responses. The unidentified jargon fits into a response template, instructing the user to check the spelling and contact the knowledge base manager to add the new term.



\section{Evaluation}

We conduct two experiments to evaluate our method's effectiveness. The first experiment tests our method's ability to answer domain-specific questions based on documents, and the second experiment tests LLM's ability to correctly identify abbreviations from questions.

\subsection{Question-Answering Experiment}

\subsubsection{Dataset Preparation}

To evaluate our method's ability to answer domain-specific questions based on documents, we collected multiple-choice questions from training documents for new-hire engineers. The questions cover six different domains, with each domain having nine to ten questions. These questions are one to two sentences long and contain jargon or abbreviations, with choices ranging from two (True/False) to four (Multiple choice). Examples of these questions are provided in Appendix \ref{Appendix: Question answering data examples}.

\subsubsection{Experiment Setup}

The questions and choices are presented to the LLM/chatbot along with instructions to select an answer. Responses are collected and graded by a human expert who records the number of correct answers for each quiz. Each quiz is repeated five times, and the average score is calculated for each method and LLM backbone.

We compare our method with vanilla LLM (without RAG) and the vanilla RAG method. For each method, including ours, we test three state-of-the-art models: Meta-Llama-3-70B-Instruct \cite{llama3modelcard}, Mixtral-8x22B-Instruct-v0.1, and Shisa-v1-Llama3-70b.2e5.

\subsubsection{Result}

We list the scores of each method and LLM backbone in Table \ref{Tab: experiment 1}. Compared with Vanilla LLM and RAG, Golden-Retriever improves the total score of Meta-Llama-3-70B by 79.2\% and 40.7\%, respectively. Across all three LLMs tested, Golden-Retriever improves the scores by an average of 57.3\% over Vanilla LLM and 35.0\% over RAG. This demonstrates that Golden-Retriever significantly enhances question-answering accuracy across multiple LLM backbones.

\subsection{Abbreviation Identification Experiment}
\begin{table}[]
\caption{Abbreviation identification accuracy.}\label{Tab: abbreviation id result}
\begin{tabular}{r|ccccc}
\toprule
Model   & \multicolumn{5}{c}{No. of Abbrev. in Question} \\
        & 1      & 2      & 3      & 4    & 5     \\
        \midrule
Llama3  & 70\%   & 100\%  & 90\%   & 90\% & 100\% \\
Mistral & 100\%  & 100\%  & 100\%  & 70\% & 80\%  \\
Shisa   & 50\%   & 80\%   & 60\%   & 80\% & 100\% \\
\bottomrule
\end{tabular}
\end{table}

\subsubsection{Dataset Preparation}

To test if LLMs can robustly identify unknown abbreviations (Section \ref{Sec: Identify Jargon}), we generated random abbreviations and inserted them into question templates to create a synthetic dataset. For abbreviation generation, we computed the probability distribution of each letter being the first letter in all words in an English dictionary, then sequentially sampled the letters by that distribution to form abbreviations. We manually prepared question templates. The question templates and generated abbreviations are shown in the random abbreviation generation code in Appendix \ref{Appendix: Random Abbreviation Generation Code}.

\subsubsection{Experiment Setup}

The synthetic questions are integrated with the prompt template, as shown in the "Identify Jargon" step in Figure \ref{Fig: Framework}. We prompt the LLM, record the responses, and check if they contain all abbreviations used in the questions. This experiment is conducted on the three aforementioned LLMs.

\subsubsection{Result}

We list the accuracy of each LLM in identifying all abbreviations in questions with varying numbers of abbreviations in Table \ref{Tab: abbreviation id result}. The experiment shows that state-of-the-art models such as Llama3 and Mistral have high accuracy in identifying unknown abbreviations. We also observe different failure modes across the three LLMs, with detailed fail cases shown in Appendix \ref{Appendix: Abbreviation identification sample experiment results}.

\section{Conclusion}

This paper presents Golden-Retriever, a novel agentic RAG system designed to efficiently navigate vast industrial knowledge bases and overcome the challenges of domain-specific jargon and context interpretation. Experiment on a dedicated question-answer dataset shows that Golden-Retriever significantly improves answer accuracy, demonstrating its superior performance compared with traditional RAG method.




\section*{Acknowledgments}

Zhiyu An would like to acknowledge Western Digital Corporation for offering generous support during the summer internship and providing the challenging problems that inspired this research.

\bibliography{custom}

\begin{thebibliography}{16}
\providecommand{\natexlab}[1]{#1}

\bibitem[{AI@Meta(2024)}]{llama3modelcard}
AI@Meta. 2024.
\newblock \href {https://github.com/meta-llama/llama3/blob/main/MODEL_CARD.md} {Llama 3 model card}.

\bibitem[{Asai et~al.(2023)Asai, Wu, Wang, Sil, and Hajishirzi}]{asai2023self}
Akari Asai, Zeqiu Wu, Yizhong Wang, Avirup Sil, and Hannaneh Hajishirzi. 2023.
\newblock Self-rag: Learning to retrieve, generate, and critique through self-reflection.
\newblock \emph{arXiv preprint arXiv:2310.11511}.

\bibitem[{Berglund et~al.(2023)Berglund, Tong, Kaufmann, Balesni, Stickland, Korbak, and Evans}]{berglund2023reversal}
Lukas Berglund, Meg Tong, Max Kaufmann, Mikita Balesni, Asa~Cooper Stickland, Tomasz Korbak, and Owain Evans. 2023.
\newblock The reversal curse: Llms trained on" a is b" fail to learn" b is a".
\newblock \emph{arXiv preprint arXiv:2309.12288}.

\bibitem[{Golovneva et~al.(2024)Golovneva, Allen-Zhu, Weston, and Sukhbaatar}]{golovneva2024reverse}
Olga Golovneva, Zeyuan Allen-Zhu, Jason Weston, and Sainbayar Sukhbaatar. 2024.
\newblock Reverse training to nurse the reversal curse.
\newblock \emph{arXiv preprint arXiv:2403.13799}.

\bibitem[{Hu et~al.(2021)Hu, Shen, Wallis, Allen-Zhu, Li, Wang, Wang, and Chen}]{hu2021lora}
Edward~J Hu, Yelong Shen, Phillip Wallis, Zeyuan Allen-Zhu, Yuanzhi Li, Shean Wang, Lu~Wang, and Weizhu Chen. 2021.
\newblock Lora: Low-rank adaptation of large language models.
\newblock \emph{arXiv preprint arXiv:2106.09685}.

\bibitem[{Kochedykov et~al.(2023)Kochedykov, Yin, and Khatravath}]{kochedykov2023conversing}
Denis Kochedykov, Fenglin Yin, and Sreevidya Khatravath. 2023.
\newblock Conversing with databases: Practical natural language querying.
\newblock In \emph{Proceedings of the 2023 Conference on Empirical Methods in Natural Language Processing: Industry Track}, pages 372--379.

\bibitem[{Lewis et~al.(2020)Lewis, Perez, Piktus, Petroni, Karpukhin, Goyal, K{\"u}ttler, Lewis, Yih, Rockt{\"a}schel et~al.}]{lewis2020retrieval}
Patrick Lewis, Ethan Perez, Aleksandra Piktus, Fabio Petroni, Vladimir Karpukhin, Naman Goyal, Heinrich K{\"u}ttler, Mike Lewis, Wen-tau Yih, Tim Rockt{\"a}schel, et~al. 2020.
\newblock Retrieval-augmented generation for knowledge-intensive nlp tasks.
\newblock \emph{Advances in Neural Information Processing Systems}, 33:9459--9474.

\bibitem[{Li et~al.(2024)Li, Hui, Qu, Yang, Li, Li, Wang, Qin, Geng, Huo et~al.}]{li2024can}
Jinyang Li, Binyuan Hui, Ge~Qu, Jiaxi Yang, Binhua Li, Bowen Li, Bailin Wang, Bowen Qin, Ruiying Geng, Nan Huo, et~al. 2024.
\newblock Can llm already serve as a database interface? a big bench for large-scale database grounded text-to-sqls.
\newblock \emph{Advances in Neural Information Processing Systems}, 36.

\bibitem[{Petroni et~al.(2019)Petroni, Rockt{\"a}schel, Lewis, Bakhtin, Wu, Miller, and Riedel}]{petroni2019language}
Fabio Petroni, Tim Rockt{\"a}schel, Patrick Lewis, Anton Bakhtin, Yuxiang Wu, Alexander~H Miller, and Sebastian Riedel. 2019.
\newblock Language models as knowledge bases?
\newblock \emph{arXiv preprint arXiv:1909.01066}.

\bibitem[{Qin et~al.(2023)Qin, Liang, Ye, Zhu, Yan, Lu, Lin, Cong, Tang, Qian et~al.}]{qin2023toolllm}
Yujia Qin, Shihao Liang, Yining Ye, Kunlun Zhu, Lan Yan, Yaxi Lu, Yankai Lin, Xin Cong, Xiangru Tang, Bill Qian, et~al. 2023.
\newblock Toolllm: Facilitating large language models to master 16000+ real-world apis.
\newblock \emph{arXiv preprint arXiv:2307.16789}.

\bibitem[{Roberts et~al.(2020)Roberts, Raffel, and Shazeer}]{roberts2020much}
Adam Roberts, Colin Raffel, and Noam Shazeer. 2020.
\newblock How much knowledge can you pack into the parameters of a language model?
\newblock \emph{arXiv preprint arXiv:2002.08910}.

\bibitem[{Sen et~al.(2023)Sen, Assenmacher, Samory, Augenstein, van~der Aalst, and Wagne}]{sen2023people}
Indira Sen, Dennis Assenmacher, Mattia Samory, Isabelle Augenstein, Wil van~der Aalst, and Claudia Wagne. 2023.
\newblock People make better edits: measuring the efficacy of llm-generated counterfactually augmented data for harmful language detection.
\newblock \emph{arXiv preprint arXiv:2311.01270}.

\bibitem[{Singhal et~al.(2023)Singhal, Azizi, Tu, Mahdavi, Wei, Chung, Scales, Tanwani, Cole-Lewis, Pfohl et~al.}]{singhal2023large}
Karan Singhal, Shekoofeh Azizi, Tao Tu, S~Sara Mahdavi, Jason Wei, Hyung~Won Chung, Nathan Scales, Ajay Tanwani, Heather Cole-Lewis, Stephen Pfohl, et~al. 2023.
\newblock Large language models encode clinical knowledge.
\newblock \emph{Nature}, 620(7972):172--180.

\bibitem[{Whitehouse et~al.(2023)Whitehouse, Choudhury, and Aji}]{whitehouse2023llm}
Chenxi Whitehouse, Monojit Choudhury, and Alham~Fikri Aji. 2023.
\newblock Llm-powered data augmentation for enhanced cross-lingual performance.
\newblock \emph{arXiv preprint arXiv:2305.14288}.

\bibitem[{Yan et~al.(2024)Yan, Gu, Zhu, and Ling}]{yan2024corrective}
Shi-Qi Yan, Jia-Chen Gu, Yun Zhu, and Zhen-Hua Ling. 2024.
\newblock Corrective retrieval augmented generation.
\newblock \emph{arXiv preprint arXiv:2401.15884}.

\bibitem[{Zhai et~al.(2024)Zhai, Tong, Li, Cai, Qu, Lee, and Ma}]{zhai2024investigating}
Yuexiang Zhai, Shengbang Tong, Xiao Li, Mu~Cai, Qing Qu, Yong~Jae Lee, and Yi~Ma. 2024.
\newblock Investigating the catastrophic forgetting in multimodal large language model fine-tuning.
\newblock In \emph{Conference on Parsimony and Learning}, pages 202--227. PMLR.

\end{thebibliography}

\newpage

\appendix

\onecolumn

\section{Fine-tuning or Retrieval Augmented Generation?}
\label{appendix: Fine-tuning or Retrieval Augmented Generation}

Knowledge injection via fine-tuning has several significant drawbacks. For instance, when fine-tuned on a knowledge statement like "A is B," the fine-tuned LLM can correctly answer "What is A?" but fails to answer "What is B?" with "A" for arbitrary A and B. This phenomenon is famously known as The Reversal Curse \citep{berglund2023reversal}. Although remedies such as generating reversed training data \citep{golovneva2024reverse} have been proposed, they require higher training costs and do not guarantee that the tuned LLM will answer all possible forms of a query. Additionally, incorporating knowledge through fine-tuning necessitates a new fine-tuning job for each new piece of knowledge, which incurs computational costs and hinders efficient integration of new information. The amount of knowledge a model can effectively incorporate depends on the capacity of the fine-tuned model part \citep{roberts2020much}, while excessive fine-tuning may lead to catastrophic forgetting, where the model forgets previously learned knowledge \citep{zhai2024investigating}.

In contrast, RAG does not suffer from these drawbacks. The Reversal Curse, observed in fine-tuning methods, does not occur when knowledge statements are presented in-context, as part of the prompt. In RAG, the LLM learns knowledge statements in-context, significantly improving its reasoning capacity and enabling efficient instruction prompt tuning \citep{singhal2023large}. Furthermore, RAG does not require model retraining and can efficiently incorporate new knowledge corpora. These properties make RAG a superior choice for industrial knowledge bases.

\section{Question-Answering Data Examples}\label{Appendix: Question answering data examples}

Here we show a few non-confidential instances of the evaluation data used in the question-answering experiment, as follows:

\noindent\fbox{%
    \parbox{\textwidth}{%
    Who decides ACT timing SPECS?\\ a. Memory Team.\\ b. System Team.\\ c. JEDEC/ONFI\\ d. Customer. \\\textbf{Answer:} c
    }%
}

\noindent\fbox{%
    \parbox{\textwidth}{%
    In any of the 2D NAND dies, CMD/ADDR protocol is of what nature?\\1. Legacy \\2. DDR1 \\3. DDR2 \\4. Depends on that particular project/technology node. \\\textbf{Answer:} 1
    }%
}

\noindent\fbox{%
    \parbox{\textwidth}{%
    In Dynamic timing analysis, timing parameters margin is checked without doing any simulations.\\a. True \\b. False. \\\textbf{Answer:} b
    }%
}

\noindent\fbox{%
    \parbox{\textwidth}{%
    What is the VCCQ level for DDR3 standard?\\1. 1.8V \\2. 1.7V \\3. 1.6V \\4. 1.5V \\5. 1.2V \\\textbf{Answer:} 4
    }%
}

\noindent\fbox{%
    \parbox{\textwidth}{%
    What is the state of ALEx,CLEx during DATA IN operation?\\a. 00\\b. 11\\c. 10\\d. Don’t care. \\\textbf{Answer:} a
    }%
}

\noindent\fbox{%
    \parbox{\textwidth}{%
    We need 5ns clock period to achieve 400MBps in DDR. To achieve 50MBps in SDR, what should be the clock period?\\1. 10ns \\2. 20ns \\3. 40ns \\4. 80ns \\\textbf{Answer:} 2
    }%
}

\noindent\fbox{%
    \parbox{\textwidth}{%
    Why do we use dummy transistors?\\a. protect the actual transistors while fabrication \\b. Can be used as spare transistors to be used in refinement of the circuit. \\c. To create uniform environment for pair transistors. \\d. All of the above \\\textbf{Answer:} d
    }%
}

\noindent\fbox{%
    \parbox{\textwidth}{%
    Which of the following tasks is not the responsibility of ACT team?\\1. IO design \\2. Data path design \\3. Pad order design \\4. Package design \\\textbf{Answer:} 4
    }%
}

\noindent\fbox{%
    \parbox{\textwidth}{%
    Which of the following factors affect Electromigration in the circuit?\\1. Number of contacts/vias at the connecting junction of two metal layers \\2. Current density in the metal layer \\3. Temperature \\4. All of above \\\textbf{Answer:} 4
    }%
}

\noindent\fbox{%
    \parbox{\textwidth}{%
    What parameter Receiver skew affects largely in the design?\\a) Input Voltage \\b) Duty Cycle \\c) Input Slew Rate \\d) Data reception \\\textbf{Answer:} b
    }%
}

\newpage

\section{Abbreviation Identification Experiment}

\subsection{Synthetic Dataset Generation Template}\label{Appendix: Random Abbreviation Generation Code}

Below is the question template and the list of random abbreviations used for generating random abbreviations in the abbreviation identification experiment.
\begin{lstlisting}[language=Python]
question_templates = [
    # Templates with one abbreviation
    "What does the abbreviation {abbr1} stand for?",
    "Can you explain the meaning of {abbr1}?",
    "What is the full form of {abbr1}?",
    "{abbr1} is an abbreviation for what?",
    
    # Templates with two abbreviations
    "What do the abbreviations {abbr1} and {abbr2} mean?",
    "In the case where {abbr1} > 0.5, how much should {abbr2} be?",
    "What is the relationship between {abbr1} and {abbr2}?",
    
    # Templates with three abbreviations
    "Consider {abbr1} = 1.5 and {abbr2} < 0.1, what would {abbr3} be?",
    "{abbr1} and {abbr2} are the same. Should {abbr3} be high or low?",
    "What is the state of {abbr1}, {abbr2} during {abbr3} operation?",
    
    # Templates with four abbreviations
    "We need 10ns {abbr1} to achieve 40{abbr2} in {abbr3}. What should be {abbr4}?",
    "In any of the {abbr1}, {abbr2}/{abbr3}/{abbr4} should be what nature?",
    
    # Template with five abbreviations
    "{abbr1}=10, {abbr2}=5, {abbr3}<0.1 in {abbr4}. How should I set {abbr5}?"
]

random_abbreviation_list = ['TS', 'IE', 'MI', 'SF', 'MP', 'UM', 'ES', 'PE', 'UW', 'SU', 'FSU', 'QMB', 'KPU', 'VMT', 'ESO', 'ARI', 'SPA', 'MTD', 'GTC', 'ODV', 'SLBG', 'MUBO', 'ROSN', 'VPPL', 'PIOF', 'CCPP', 'MBST', 'UTUU', 'NIRE', 'STUP']
\end{lstlisting}

\subsection{Sample Experiment Results}\label{Appendix: Abbreviation identification sample experiment results}

Here we present the sample experiment results from the abbreviation identification experiment. We emphasize on demonstrating the failure modes, which is different across LLMs.

\subsubsection{Meta Llama 3}

\begin{lstlisting}[language=Python]
[
    {
        "Question": "What is the full form of KPU?",
        "Response": "[\"KPU\"]" # correct
    },
    {
        "Question": "What is the full form of ARI?",
        "Response": "[\"ARI\"]" # correct
    },
    {
        "Question": "Can you explain the meaning of GTC?",
        "Response": "[\"GTC\", \"Good Till Cancelled\"]" # incorrect
    },
    {
        "Question": "Can you explain the meaning of SPA?",
        "Response": "[\"Single-Page Application\", \"SPA\"]" # incorrect
    },
    {
        "Question": "In any of the UTUU, ES/NIRE/MUBO should be what nature?",
        "Response": "[\"UTUU\", \"ES/NIRE/MUBO\"]" # incorrect
    },
]
\end{lstlisting}

\subsubsection{Mistral}

\begin{lstlisting}[language=Python]
[
    {
        "Question": "We need 10ns ARI to achieve 40MI in MUBO. What should be PIOF?",
        "Response": "[\"ARI\", \"MI\", \"MUBO\", \"PIOF\"]" # correct
    },
    {
        "Question": "We need 10ns NIRE to achieve 40MP in STUP. What should be IE?",
        "Response": "[\"10ns NIRE\", \"40MP in STUP\", \"IE\"]" #incorrect
    },
    {
        "Question": "We need 10ns MBST to achieve 40ROSN in SPA. What should be UW?",
        "Response": "[\"10ns MBST\", \"40ROSN\", \"SPA\", \"UW\"]" # incorrect
    },
    {
        "Question": "In any of the UTUU, ES/NIRE/MUBO should be what nature?",
        "Response": "[\"UTUU\", \"ES/NIRE/MUBO\"]" # incorrect
    },
]
\end{lstlisting}

\subsubsection{Shisa}

\begin{lstlisting}[language=Python]
[
    {
        "Question": "PIOF is an abbreviation for what?",
        "Response": "[\"PIOF\"]</s>" # correct
    },
    {
        "Question": "What is the full form of IE?",
        "Response": "[\"IE\", \"Internet Explorer\"]\n\nPlease let me know if you need anything else.</s>" # incorrect
    },
    {
        "Question": "SPA is an abbreviation for what?",
        "Response": "[\"SPA\"]\n\nPlease let me know if I can assist you further.</s>" # incorrect
    },
    {
        "Question": "Can you explain the meaning of SPA?",
        "Response": "[\"SPA\"]\n\nThe text translated to Japanese is:\n[\u300cSPA\u300d]</s>" # incorrect
    },
    {
        "Question": "Can you explain the meaning of SU?",
        "Response": "[\"SU\"]\n\nThe text translated to Japanese is:\n### \u6307\u793a:\n\u300cSU\u300d\u3068\u3044\u3046\u610f\u5473\u3092\u8aac\u660e\u3067\u304d\u307e\u3059\u304b\uff1f\u3053\u306e\u8cea\u554f\u304b\u3089\u3001\u5c02\u9580\u7528\u8a9e\u3084\u7565\u8a9e\u3092\u7279\u5b9a\u3057\u3066\u304f\u3060\u3055\u3044\u3002\u6b21\u306e\u5f62\u5f0f\u3067\u30ea\u30b9\u30c8\u30a2\u30c3\u30d7\u3057\u3066\u304f\u3060\u3055\u3044: [\u300c\u5c02\u9580\u7528\u8a9e1\u300d\u3001\u300c\u5c02\u9580\u7528\u8a9e2\u300d\u3001...]\n\n### \u56de\u7b54:\n[\"SU\"]</s>" # incorrect
    },
    {
        "Question": "What is the relationship between SU and SF?",
        "Response": "[\u201cSU\u201d, \u201cSF\u201d]</s>" # incorrect
    },
]
\end{lstlisting}

\end{document}